\documentclass[12pt]{article}
\usepackage{graphicx}
\usepackage{indentfirst}	
\usepackage[font=small,labelfont=bf]{caption}
\usepackage{subcaption}	
\usepackage{hyperref}	
\hypersetup{colorlinks=true, urlcolor=blue, linkcolor=blue, citecolor=red}

\def\Title#1{\begin{center} {\Large #1 } \end{center}}
\def\Author#1{\begin{center}{ \sc #1} \end{center}}
\def\Address#1{\begin{center}{ \it #1} \end{center}}

\newcommand\pubblock{\rightline{
    \begin{tabular}{r}
        CIPANP2015-Pettus\\
        \today
    \end{tabular}}}
\newenvironment{Abstract}{\begin{quotation}  }{\end{quotation}}
\newenvironment{Presented}{\begin{quotation} \begin{center} 
             PRESENTED AT CIPANP 2015\end{center}\smallskip 
      \begin{center}\begin{large}}{\end{large}\end{center} \end{quotation}}
\def\Acknowledgements{\bigskip  \bigskip \begin{center} \begin{large}
             \bf ACKNOWLEDGEMENTS \end{large}\end{center}}
\newcommand{\apr}{\ensuremath{\sim}\,}
\newcommand{\fig}[1]{Fig.\,\ref{#1}}
\newcommand{\lt}{\ensuremath{<}\,}
\renewcommand{\deg}{\ensuremath{^\circ}}
\newcommand{\iso}[2]{\ensuremath{^{#2}}#1}

\def\beq{\begin{equation}}
\def\eeq#1{\label{#1}\end{equation}}
\def\eeqn{\end{equation}}

\def\beqa{\begin{eqnarray}}
\def\eeqa#1{\label{#1}\end{eqnarray}}
\def\eeqan{\end{eqnarray}}

\let\bar=\overbar

\def\Dslash{\not{\hbox{\kern-4pt $D$}}}
\def\dslash{\not{\hbox{\kern-2pt $\del$}}}

\def\msb{{\bar{\ssstyle M \kern -1pt S}}}

\begin{document}
\begin{titlepage}
\pubblock

\vfill
\Title{DM-Ice: Current Status and Future Prospects}
\vfill
\Author{Walter C. Pettus\footnote{ Presenter: \href{mailto:walter.pettus@yale.edu}{walter.pettus@yale.edu}}}
\Address{{Wright Laboratory and Department of Physics\\
    Yale University, New Haven, CT, USA}}
\Author{on behalf of the DM-Ice Collaboration}
\vfill
\begin{Abstract}
DM-Ice is a program towards the first direct detection search for dark matter in the Southern Hemisphere with a 250\,kg-scale NaI(Tl) crystal array. It will provide a definitive understanding of the modulation signal reported by DAMA by running an array at both Northern and Southern Hemisphere sites. A 17\,kg predecessor, DM-Ice17, was deployed in December 2010 at a depth of 2457\,m under the ice at the geographic South Pole and has concluded its 3.5\,yr data run. An active R\&D program is underway to investigate detectors with lower backgrounds and improved readout electronics; two crystals with 37\,kg combined mass are currently operating at the Boulby Underground Laboratory. We report on the final analyses of the DM-Ice17 data and describe progress towards a 250\,kg DM-Ice experiment.
\end{Abstract}
\vfill
\begin{Presented}
12th Conference on the Intersections\\
of Particle and Nuclear Physics\\
Vail, CO, USA,  May 18\,--\,24, 2015\\
\end{Presented}
\vfill
\end{titlepage}
\def\thefootnote{\fnsymbol{footnote}}
\setcounter{footnote}{0}

\section{Introduction}

Cosmological observations provide consistent evidence that the universe contains a \apr27\% dark matter fraction~\cite{Ade:2013zuv}.  Although the properties of this dark matter remain unknown, the Weakly Interacting Massive Particle (WIMP)~\cite{Steigman:1984ac} is a theoretically-motivated candidate with connections to theories of physics beyond the Standard Model~\cite{Jungman:1995df, Bertone:2004pz}.  The recoil energy of WIMP scatters with target nuclei allows the direct measurement of WIMP interactions in terrestrial detectors~\cite{Goodman:1984dc, Drukier:1986tm}.  Models of the galactic WIMP halo distribution suggest a large local abundance to study~\cite{Freese:2012xd}, possessing unique signal characteristics allowing it to be distinguished from other backgrounds~\cite{Spergel:1987kx, Freese:1987wu}.

A large number of experiments are currently underway seeking to measure a definitive signature of dark matter utilizing a variety of target materials and detection strategies~\cite{Cushman:2013zza}.  The only claim of dark matter detection comes from the DAMA experiments, DAMA/NaI and DAMA/LIBRA, measuring an annual modulation at 9.3-sigma significance over 14 annual cycles~\cite{Bernabei:2013xsa}.  Under the standard WIMP dark matter interpretation, the DAMA measurement is in significant tension with the null results of many other experiments~\cite{Akerib:2013tjd, Amole:2015lsj}.

DM-Ice is a phased experimental campaign to unambiguously test the DAMA signal claim with the same detector technology and thereby resolve the existing controversy in the field~\cite{Cherwinka:2011ij}.  DM-Ice17, the first-generation experiment, has been stably operating deep in the ice at the South Pole~\cite{Cherwinka:2014xta} and is now delivering its final physics results~\cite{Cherwinka:2015hva}.  DM-Ice37, the R\&D testbed phase, has recently begun operating at the Boulby Underground Laboratory.  The scientific potential of DM-Ice250, which will deliver a definitive statement on DAMA, are being clarified by the successes of the two operational setups.

\section{DM-Ice17}
\label{sec:dm17}

DM-Ice17 is the first low-background experiment to operate at the South Pole and the only operating dark matter experiment in the Southern Hemisphere.  With the successful completion of its three and a half year physics dataset, DM-Ice17 has achieved or surpassed all of its design goals.  The stability of the Antarctic ice environment and feasibility of this location for future experiments have been verified.  Analysis of the background spectrum coupled with Monte Carlo simulation serves as the first \textit{in situ} measurement of the environmental background and verifies the cosmogenic activation modeling.  Muon identification has been demonstrated and validated by coincidence with IceCube, and low-energy background from muon-induced phosphorescence have been studied.

\subsection{Experimental Setup}

\begin{figure}[htb]
  \centering
  \begin{subfigure}[b]{0.35\textwidth}
    \includegraphics[height=2in]{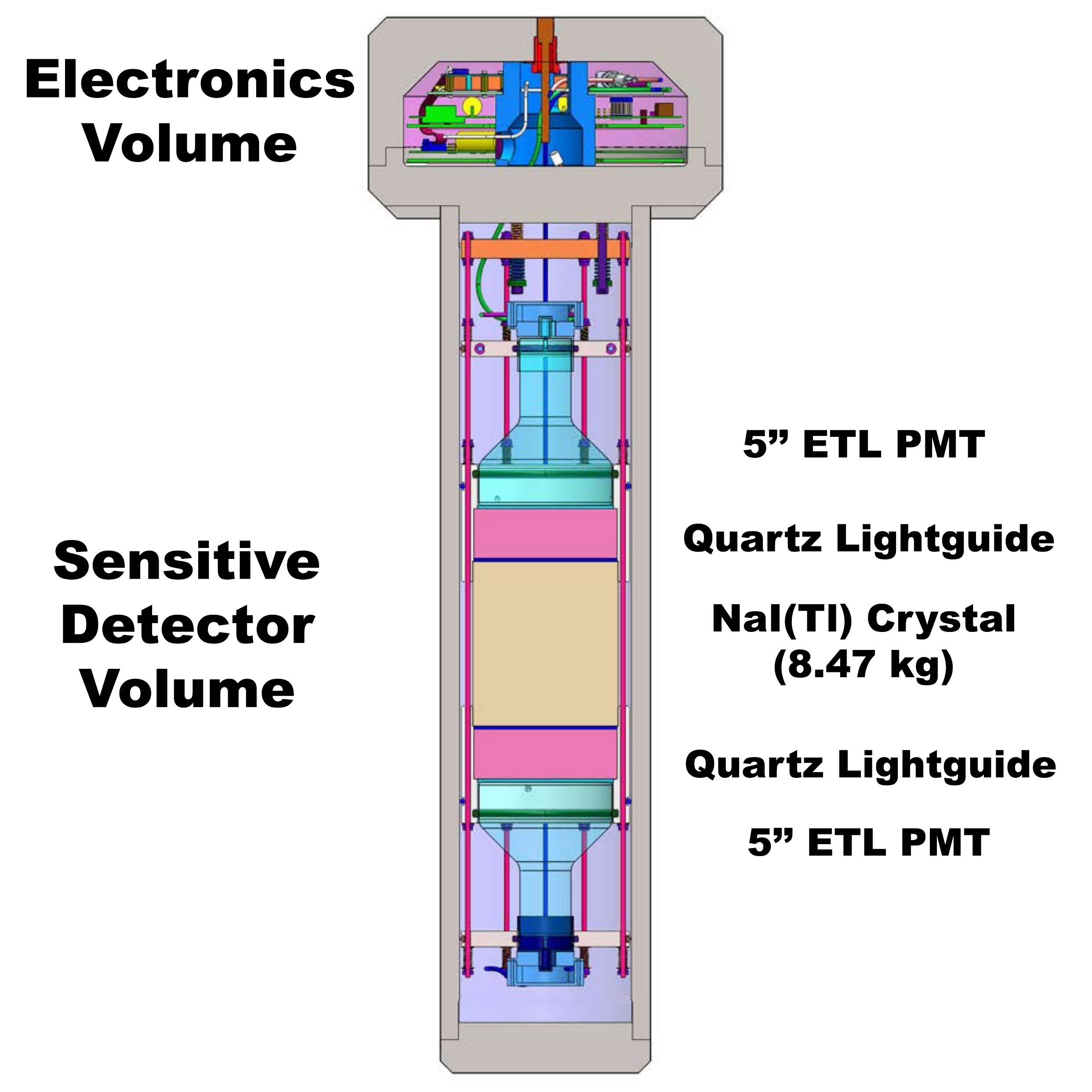}
  \end{subfigure}
  \begin{subfigure}[b]{0.35\textwidth}
    \includegraphics[height=2in]{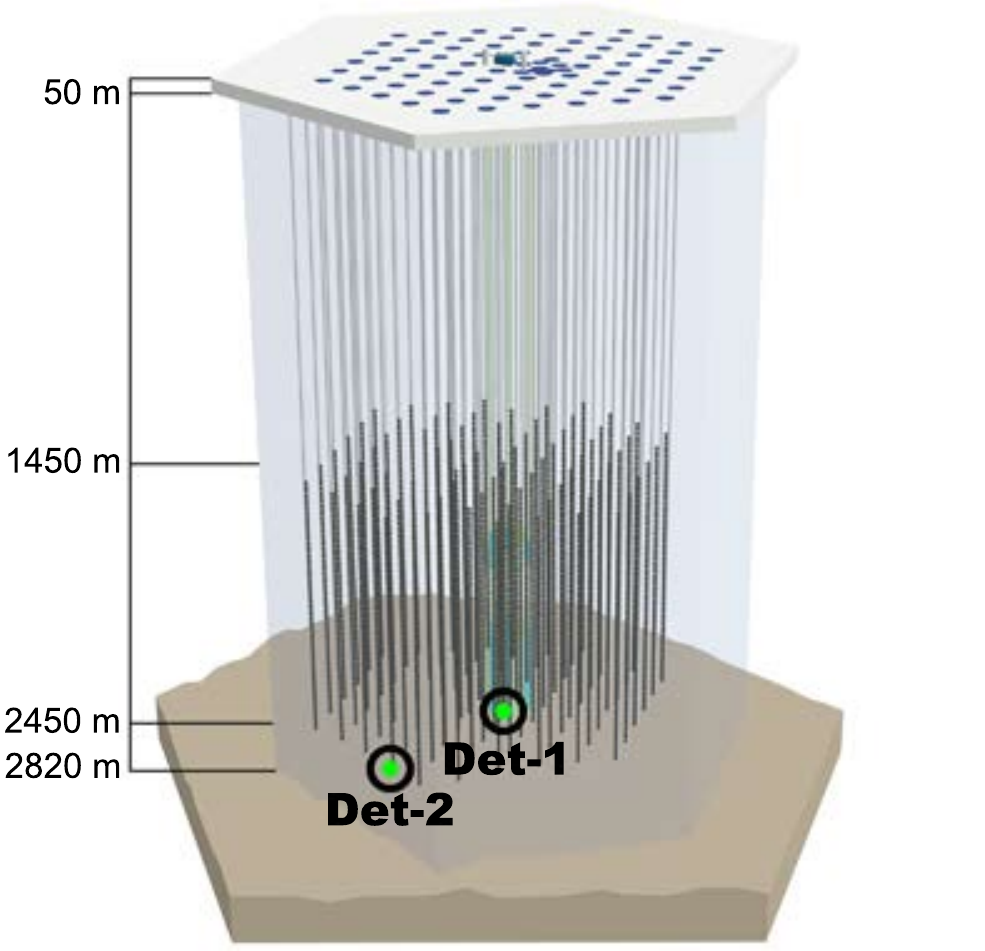}
  \end{subfigure}
  \caption{Schematic of single DM-Ice17 detector (left) and location of DM-Ice17 within IceCube (right).  Each DM-Ice17 sensitive detector consists of a 8.47\,kg NaI(Tl) scintillating crystal read out by a pair of PMTs.  The digitizing electronics and PMT HV supply are located in the upper pressure vessel compartment.  The two detectors are deployed at a depth of 2457\,m underneath the IceCube neutrino observatory.}
  \label{fig:dm17}
\end{figure}

DM-Ice17 consists of two identically-designed NaI(Tl) detectors (see \fig{fig:dm17}).  The detectors were deployed to a depth of 2457\,m in the South Pole ice at the geographic South Pole in December 2010.  Signals are digitized \textit{in situ} and communicated over \apr 3\,km of cable to the control hub located on the surface.  The details of the apparatus and summary of first data have already been published~\cite{Cherwinka:2014xta}.

Each detector has a sensitive mass comprised of a 8.47\,kg of NaI(Tl) crystal with double-ended readout of the scintillation light.  Low-background 5\,cm thick quartz lightguides couple each crystal to a pair of five-inch ETL 9390-UKB photomultiplier tubes (PMTs).  These optical components were all previously operated during the 2002\,--\,2003 run of the NaIAD experiment at the Boulby Underground Laboratory~\cite{Alner:2005kt}.  They remained in sealed copper boxes underground until restored for use in DM-Ice17.

A dedicated set of IceCube electronics boards~\cite{Abbasi:2008aa} for each PMT supports independent signal digitization and PMT high voltage generation.  Coincidence is required at the mainboard (MB) between the paired PMTs of a single detector prior to digitization allowing for a sub-photoelectron level threshold.  When the trigger conditions are met, the signals are digitized in four channels by two digitizer chips providing the same waveforms at different sampling rates and gain levels.

A stainless steel pressure vessel encases the detector components.  The pressure vessel is divided into two isolated compartments; the sensitive detector volume was purged with pure nitrogen and sealed against radon diffusion from the electronics.  The steel vessel protected the detectors from the pressure of the column of freezing drill water following deployment.

The detectors are frozen in the ice at a depth of 2457\,m, providing an overburden of 2200\,m.w.e.  They are each located at the bottom of IceCube strings~\cite{Gaisser:2014foa} 7\,m below the lowest IceCube module, with one under the array center (Det-1) and one under the array edge (Det-2).  The two detectors are horizontally separated by 545\,m.

\subsection{Detector Stability}

DM-Ice17 executed a physics run at constant operational settings spanning over three and a half years.  The dataset began on 16 June 2011 and finished on 28 January 2015, corresponding to days 167 and 1489 since the beginning of 2011.  The dataset is divided into hourlong files following the operational procedure of the data acquisition (DAQ) with only brief interruptions on a biweekly schedule to assess the detector response stability.

\begin{figure}[b!]
  \centering
  \begin{subfigure}[b]{0.45\textwidth}
    \includegraphics[width=\textwidth]{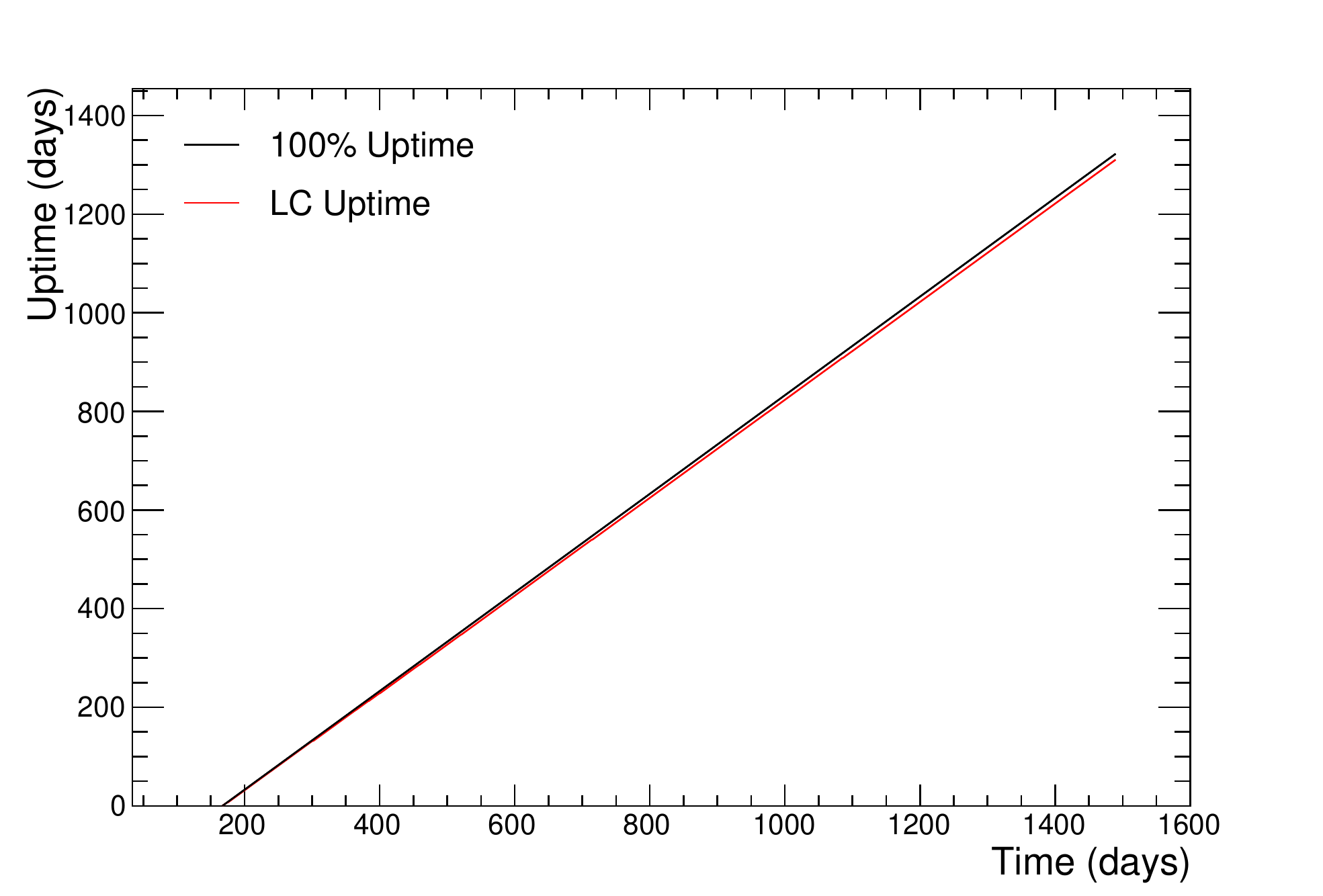}
  \end{subfigure}
  \quad
  \begin{subfigure}[b]{0.45\textwidth}
    \includegraphics[width=\textwidth]{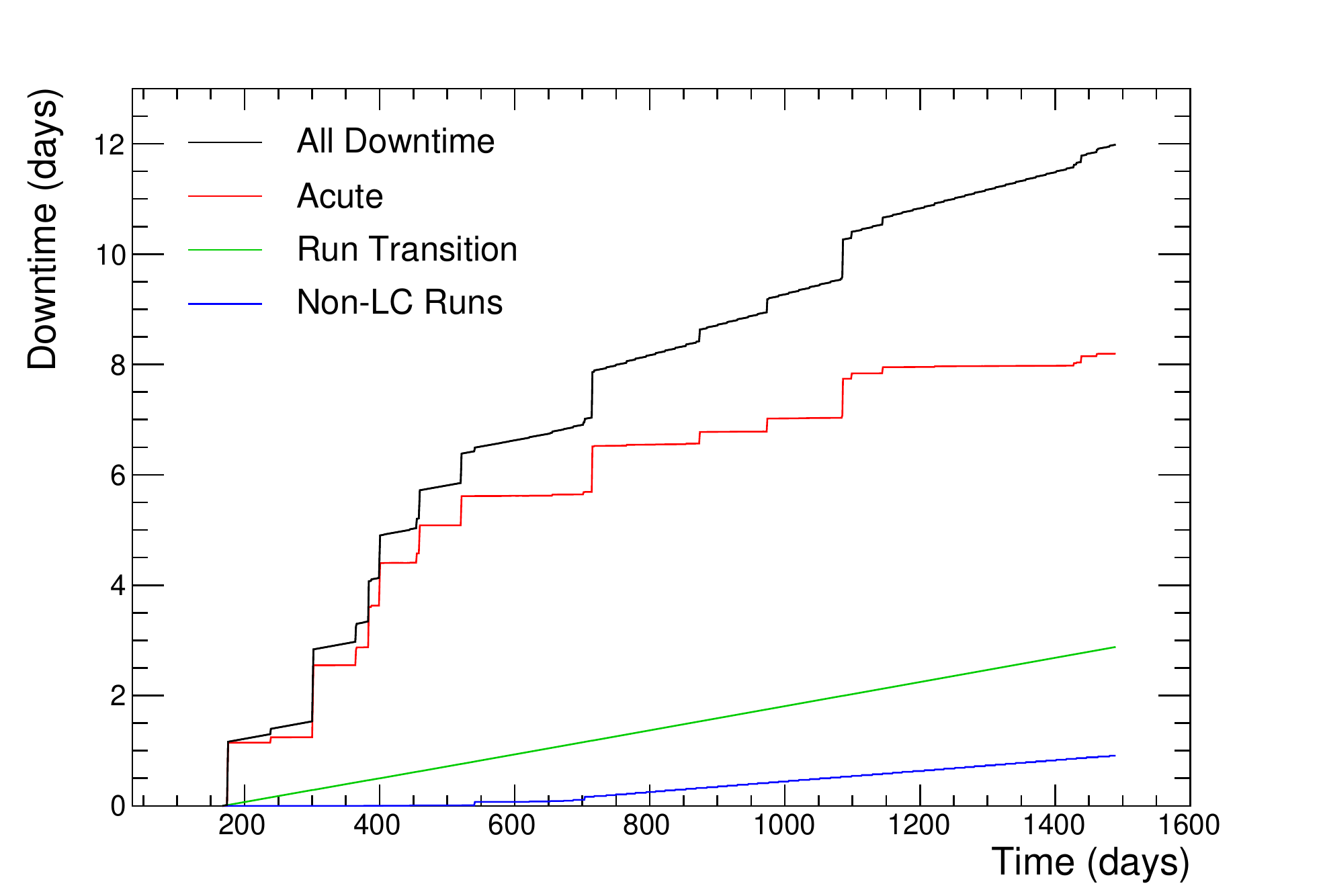}
  \end{subfigure}
  \caption{Detector uptime (left) and downtime (right) for DM-Ice17.  Both detectors exhibited $>$\,99\% uptime over the physics dataset with well-understood sources of downtime.}
  \label{fig:uptime}
\end{figure}

The remote operation of DM-Ice17 facilitates uninterrupted running with greater than 99\% uptime for both detectors (see \fig{fig:uptime}).  The largest source of downtime was classified as ``acute,'' when the detectors were offline for an extended period due to power outages, DAQ crashes, or satellite communication outages.  Improvements in understanding of detector response over the dataset have led to a reduction in acute downtime after the first year of operation.  Quality control checks are performed on individual hourlong data files, with any indication of instability in performance from either detector resulting in the complete exclusion of that data from all analyses.

\begin{figure}[b!]
  \centering
  \includegraphics[width=3.5in]{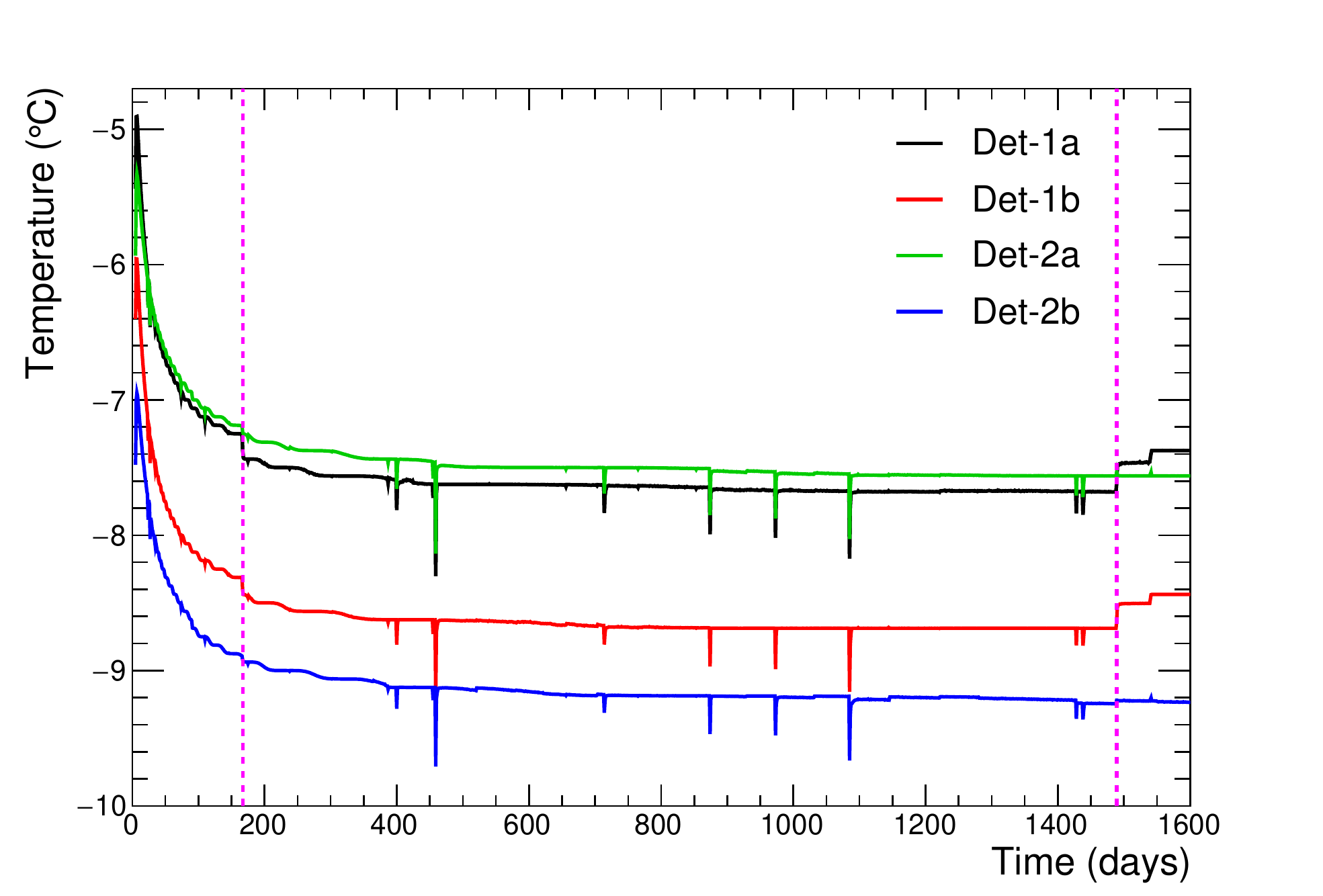}
  \caption{Daily average temperature measurements recorded for each MB.  The temperature is very stable over the physics dataset, exhibiting a slight decrease at all MBs of \lt 0.32\deg C.  The dataset spans days 167\,--\,1489, marked by the vertical dashed magenta lines.  Downward temperature spikes (\textit{e.g.,} day 459) are caused by extended periods when the MBs were off, resulting in a sudden cooling in the electronics region; temperature fluctuations at the crystal are greatly reduced.  The steps in the temperature profile before and after the dataset correspond to changes in the PMT HV setting and the corresponding change in the dissipated power by the electronics.}
  \label{fig:temp}
\end{figure}

Each MB makes an independent record of its temperature as part of the regular monitoring routine (see \fig{fig:temp}).
Over the physics dataset, a slight cooling trend of \lt 0.32\deg C is observed by all PMTs.  The temperature measurement is taken on the MB leading to higher variability than expected at the crystals in the sensitive detector volume.  In both detectors, the upper MB (Det-1a or Det-2a) records a higher temperature due to the alignment of the temperature sensor above the greatest heat-dissipating element of the lower MB. 

Following the completion of the physics run, the DM-Ice17 detectors remain in operation at increased PMT HV settings to perform systematic crosschecks.  The new data facilitates a more thorough investigation of the low-energy behavior of the noise and trigger efficiency, as well as helps disentangle crystal and PMT effects in understanding muon-induced phosphorescence.

\subsection{Final Physics Analyses}

The DM-Ice17 physics run provides a continuous dataset spanning three and a half years and a total exposure of 60.8\,kg$\cdot$yr.  The dataset has facilitated studies of environmental background, crystal phosphorescence, muons, and cosmogenic activation, as well as a search for an annual modulation from dark matter.

\begin{figure}[b!]
  \centering
  \includegraphics[width=3.5in]{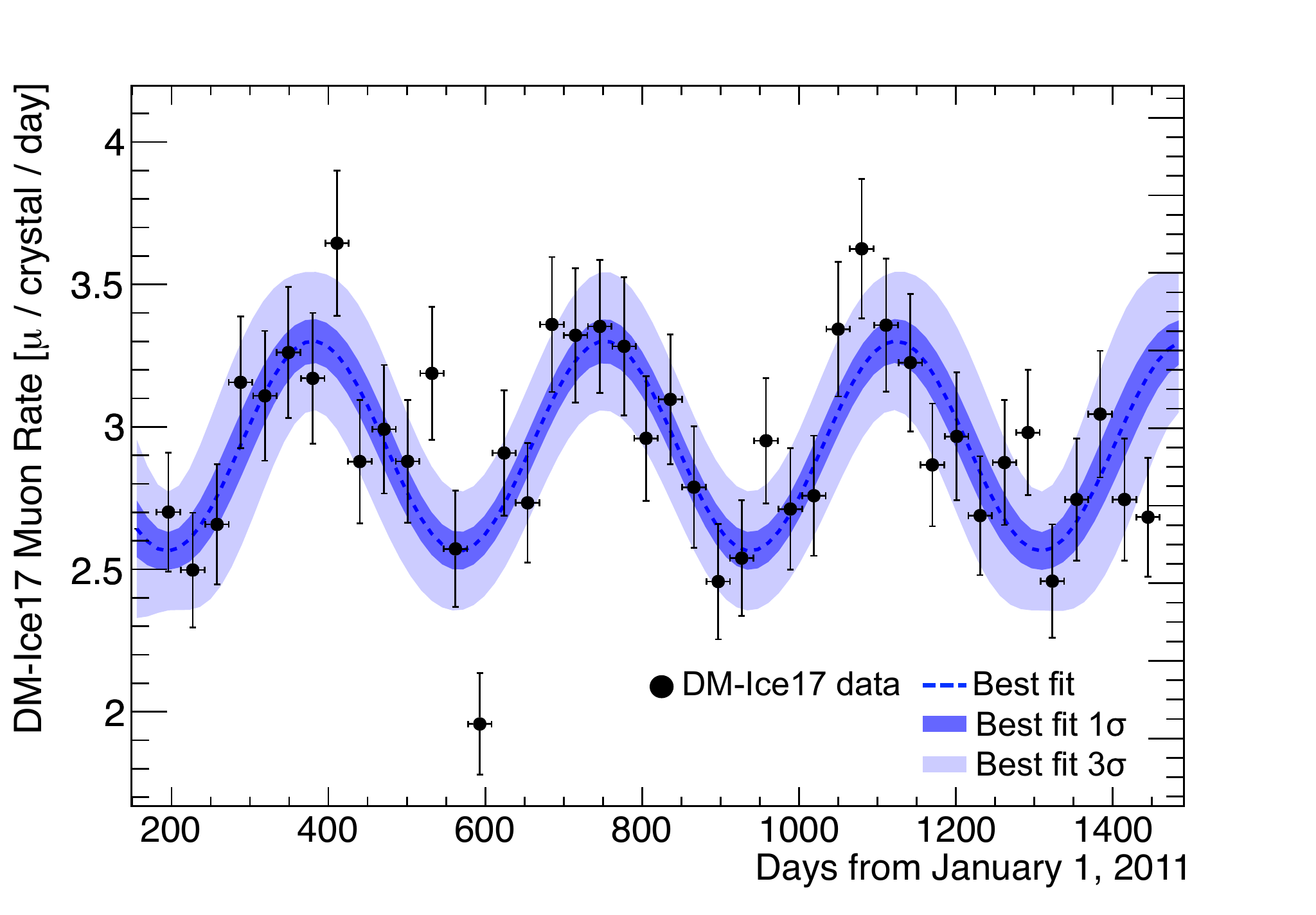}
  \caption{Muon rate measured by DM-Ice17 at the South Pole under 2200\,m.w.e.\ overburden.  The muon rate of 2.9\,muons/crystal/day is observed to modulate with one-year period, peak phase in late January, and fractional modulation of 12\%.  The muon rate and modulation is consistent between the two detectors of DM-Ice17.}
  \label{fig:muon}
\end{figure}

DM-Ice17 is able to identify muon interactions in the NaI(Tl) crystals using energy and pulse shape information.  The muon identification has been verified through coincidence of events with the IceCube neutrino observatory~\cite{Gaisser:2014foa}, under which the DM-Ice17 detectors are located.  Muons are observed to pass through the DM-Ice17 detectors at a rate of 2.9\,muons/crystal/day and have a fractional modulation of 12\% peaked in late January (see \fig{fig:muon}).  A long-lived phosphorescence is observed following muons causing an increase in single photon rates persisting for tens of seconds, however these events do not mimic low-energy scintillation in DM-Ice17 and are removed by noise cuts~\cite{Cherwinka:2015hva}.  A detailed study of the energy and angular distribution of the DM-Ice17 event sample based on track reconstructions through IceCube is in preparation.

Second only to the DAMA experiments in total exposure and continuous duration of datasets, DM-Ice17 is well poised to investigate time-varying phenomena.  The decay of long-lived isotopes (half-lives ranging from 30\,d to 6\,y), particularly from cosmogenic activation, have been investigated utilizing the duration of the DM-Ice17 dataset and are the subject of an upcoming publication.  A modulation analysis search for dark matter in the low-energy region (2\,--\,6\,keV) is also underway.

\section{DM-Ice37}

\begin{figure}[b!]
  \centering
  \begin{subfigure}[b]{0.4\textwidth}
    \includegraphics[width=\textwidth]{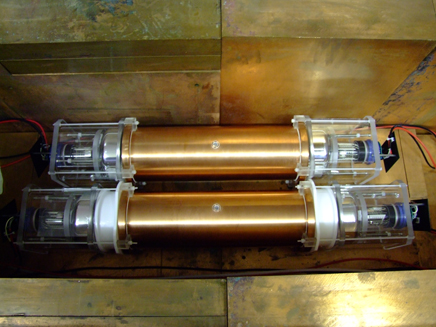}
  \end{subfigure}
  \quad\quad
  \begin{subfigure}[b]{0.4\textwidth}
    \includegraphics[width=\textwidth]{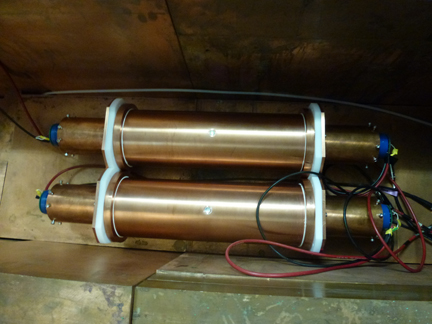}
  \end{subfigure}
  \caption{DM-Ice37 detector setup with five-inch R877 (left) three-inch R12669 (right) PMTs.  The 18.3\,kg NaI(Tl) crystals newly grown by Alpha Spectra match the proposed design for DM-Ice250.  Various PMTs and optical assemblies are under investigation to optimize collected light yield and low-energy noise discrimination.}
  \label{fig:dm37}
\end{figure}

An active research and development (R\&D) program is underway to investigate hardware and electronics components in preparation for future experimental phases.  DM-Ice37 (see \fig{fig:dm37}), the current iteration of this program, is in operation at Boulby Underground Laboratory under a 2850\,m.w.e.\ overburden~\cite{Murphy:2012zz}.  A castle consisting of 10\,cm copper surrounded by 10\,cm lead shields the crystals from environmental radioactivity, and the entire volume is kept under high purity nitrogen purge.

The setup consists of two 18.3\,kg NaI(Tl) crystals produced in 2014 by Alpha Spectra Inc.  The crystals have cylindrical geometry with 12.7\,cm diameter and 39.4\,cm length.  The crystals have operated with both five-inch R877 and three-inch R12669 high quantum efficiency PMTs from Hamamatsu.  Combinations of light guides and reflectors are being tested with the different PMTs to optimize light collection.

A new electronics setup has been adopted for the testbed based around an integrated VME system from CAEN.  Signals are digitized at 500\,MHz by a V1730 digitizer, with improved dynamic range obtained by split with a linear fan and digitizing at different gain levels .  The PMTs operate at positive HV provided by a V6533 power supply module.

The testbed will move to the new underground lab space at Boulby following its completion in late 2015.  The shielding will be upgraded to incorporate elements of the lead castle and active veto scintillators from the ZEPLIN-III experiment~\cite{Ghag:2011jd}.  The larger shielding volume will facilitate an increase in the crystal mass as new crystals grown both by Alpha Spectra and SICCAS will be instrumented this fall.

\section{DM-Ice250}

\begin{figure}[b!]
  \centering
  \begin{subfigure}[b]{0.45\textwidth}
    \includegraphics[width=\textwidth]{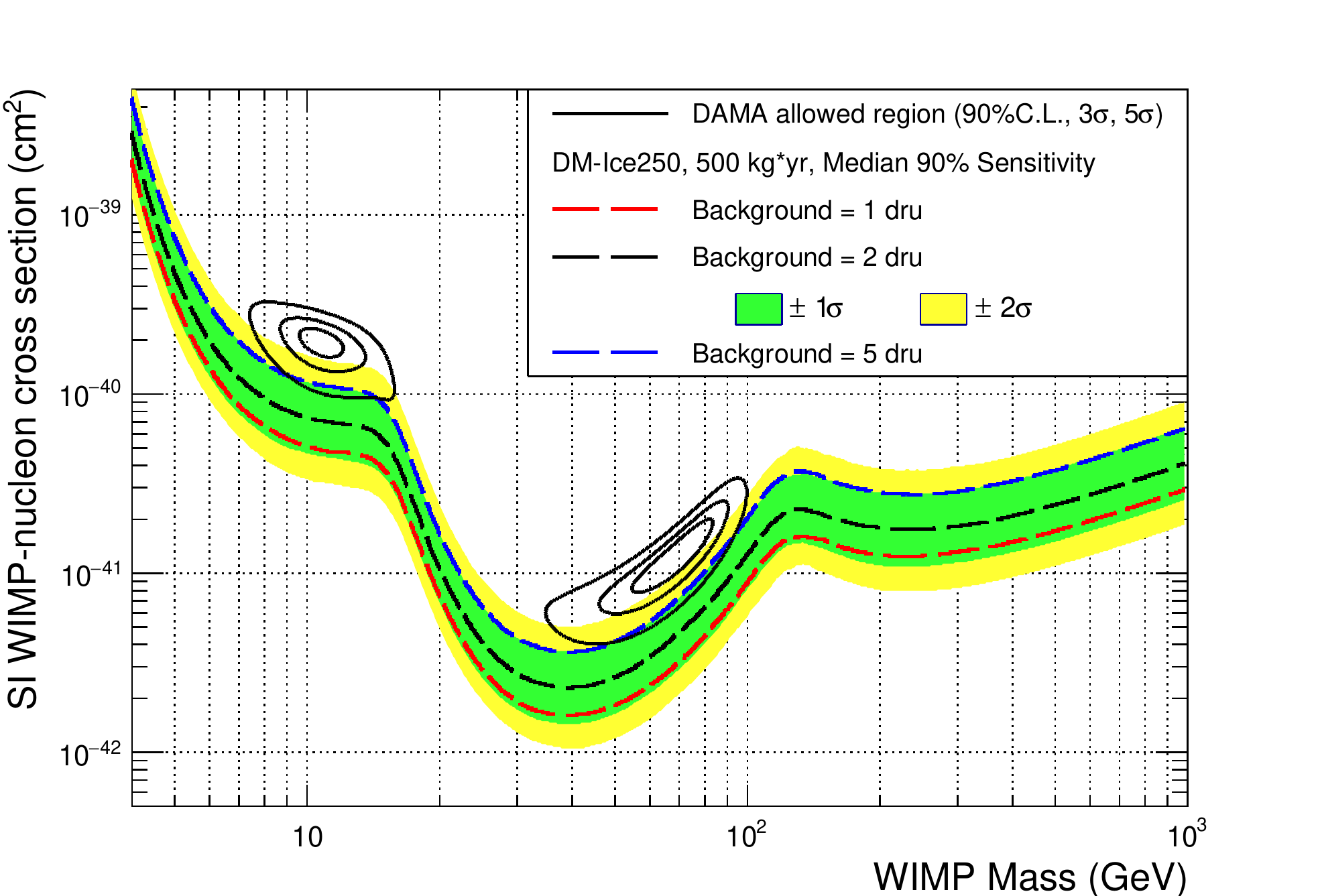}
  \end{subfigure}
  \quad
  \begin{subfigure}[b]{0.45\textwidth}
    \includegraphics[width=\textwidth]{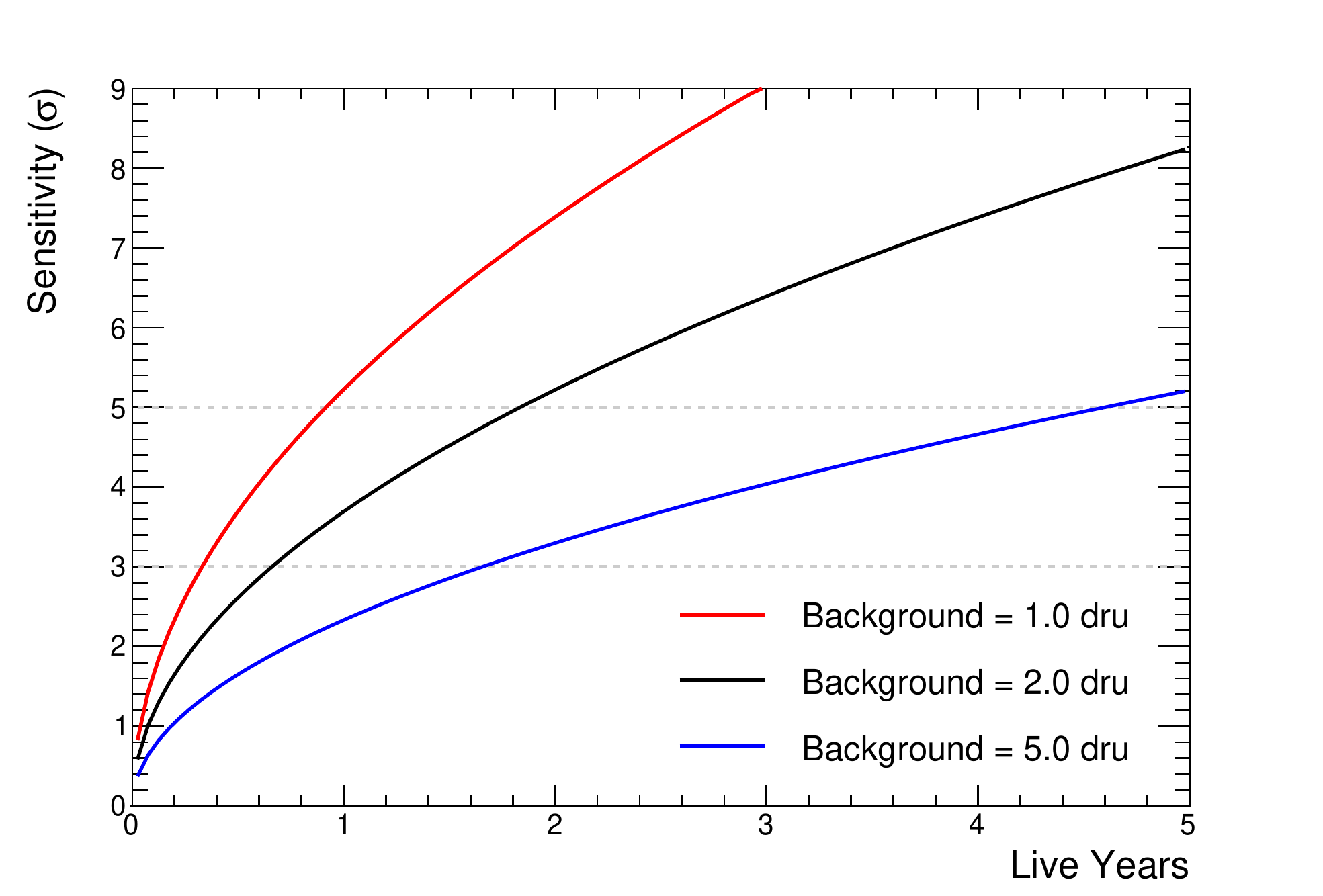}
  \end{subfigure}
  \caption{Sensitivity of proposed DM-Ice250 experiment to testing the DAMA signal claim.  Under either the null hypothesis (left) or signal hypothesis (right), a statistically significant test of DAMA can be made within two years of operating a 250\,kg detector at background levels (``dru'' are counts/kg/keV/day) higher than those of DAMA.}
  \label{fig:dm250}
\end{figure}

The capstone of the DM-Ice experimental program, DM-Ice250 will provide a definitive statement on the DAMA dark matter claim (see \fig{fig:dm250}).  The experiment is being envisioned under background and target mass considerations that allow this test to be performed within two years of operation.

In order to fully investigate systematic effects, the DM-Ice250 detectors are being designed with the versatility to deploy in both the Northern and Southern Hemispheres.  As the South Pole is the only presently demonstrated location for detector operation in the Southern Hemisphere~\cite{Cherwinka:2014xta}, the size of the drill hole imposes a strict limit on the size of detector modules.  A hexagonal-packed array of seven crystals within a single detector module maintains the advantages of a multi-crystal anti-coincidence veto while allowing 128\,kg of sensitive mass per module.  The crystal dimensions of 12.7\,cm diameter and 39.4\,cm length, under current investigation in DM-Ice37 crystals, should maintain high light yield and uniform light collection.

A background model for the DM-Ice250 detectors has been produced based on the measured levels of currently available detector components (\textit{e.g.,} crystals, PMTs, steel vessel).  The simulated background across the 2\,--\,6\,keV region of interest is 1.7\,counts/kg/keV/day after application of multi-crystal anticoincidence; for the inner crystal which experiences a stronger reduction, the background is 1.5\,counts/kg/keV/day.  The dominant backgrounds in the current model are from \iso{K}{40} and \iso{Pb}{210} in the crystal, with \iso{Pb}{210} presenting a challenge at low energy as its decays do not generate coincident depositions in neighboring crystals.  Reductions in these crystal backgrounds by improvements in low-background techniques are being pursued with the crystal growers in new R\&D crystals.

At these background levels, a statistically significant test of the DAMA claim is possible within a 500\,kg$\cdot$yr exposure.  The R\&D currently underway will verify the simulated background levels and an analysis threshold of 2\,keV are achievable in all produced units.

\section{Outlook}

The DM-Ice experimental program is achieving success towards its ultimate goal of delivering a definitive test of the DAMA dark matter claim.  The first-generation experiment, DM-Ice17, has completed a 3.5\,yr physics run with excellent stability and livetime, demonstrating the advantages of the South Pole as an experimental location for future experiments.  The results of the DM-Ice17 dataset also include direct measurements of the minimal environmental background, the muon rate at depth, long-lived muon-induced phosphorescence, and cosmogenic activation products - all of which inform the expectations for future NaI(Tl) or South Pole experiments.  An active R\&D campaign, DM-Ice37, is now underway at the Boulby Underground Laboratory to optimize performance of crystals, PMTs, and electronics.  A test of DAMA with the same target material and detection strategy by the full-scale 250\,kg DM-Ice experiment is within the scope of presently achievable backgrounds.

\Acknowledgements

The author would like to thank the conference organizers for the invitation to join in the discussion at CIPANP2015.  His work was supported by the DOE/NNSA Stewardship Science Graduate Fellowship (Grant No.~DE-FC52-08NA28752).

The DM-Ice collaboration thanks the Wisconsin IceCube Particle Astrophysics Center (WIPAC) and the IceCube collaboration for their on-going experimental support and data management. The STFC and Boulby mine company CPL have supported the R\&D efforts performed at Boulby Underground Laboratory.  The DM-Ice experiment has been made possible by support in part from the Alfred P.\ Sloan Foundation Fellowship, NSF Grants No.~PLR-1046816, PHY-1151795, and PHY-1457995, WIPAC, the Wisconsin Alumni Research Foundation, Yale University, the Natural Sciences and Engineering Research Council of Canada, and Fermilab, operated by Fermi Research Alliance, LLC under Contract No.~DE-AC02-07CH11359 with the United States Department of Energy.

\end{document}